# Designing a brown planthoppers surveillance network based on wireless sensor network approach*

Hoai Bao Lam, Tai Tan Phan, Long Huynh Vuong, Hiep Xuan Huynh, Bernard Pottier

*Abstract* - This paper proposes a new approach for monitoring brown planthoppers (BPH) swarms using a surveillance network at provincial scale. The topology of this network is identified to a *wireless sensor network* (WSN), where each node is a real light trap and each edge describes the influence between two nodes, allowing gathering BPH information. Different communication ranges are evaluated to choose a suitable network. The experiments are performed on the light traps surveillance network of Hau Giang province, a typical rice province in the Mekong Delta region of Vietnam.

## I. INTRODUCTION

The brown planthopper is a major insect pest of rice in Mekong Delta, southern Vietnam. From 2005-2006, the outbreak of BPH occurred and two virus diseases transmitted by BPH, Rice Ragged Stunt Virus disease (RRSV) and Rice Grassy Stunt Virus (RGSV) disease [11], spread over in the delta, resulting in big loss of rice production. Indeed, in that period of time, rice production in Vietnam, particularly in the Mekong Delta, suffered a major setback when outbreaks of BPH caused a loss of ~400,000 tons (1.1% of national production) [12]. Therefore, it is necessary to properly control the highly-virulent BPH.

To forecast the population of BPH, a sampling distribution with more than 340 light traps has been developed in the Mekong Delta since 2005 [1]. By monitoring the light traps, farmers will know better what types of insect are in their fields and if they are in a controllable level, or not [17].

A wireless sensor network consists of spatially distributed autonomous sensors to monitor physical or environmental conditions, such as temperature, sound, pressure, and to cooperatively pass their data through the network to a main location [4]. WSN consists of small systems, where each controller is connected to one, or sometimes several sensors.

In this paper, we consider BPH light traps networks from a WSN point of view. We can take several advantages from the WSN. One problem is how we can identify the density of BPH in a light trap location? In Mekong Delta, people usually do this task by hand, a tedious and time consuming task. Why don't we use sensors to do that? This brings the idea that each light trap could contain sensors to monitor BPH behaviors. Therefore, the network becomes a massive coordinated sensing machine. Currently, the topology of a light traps network is similar to mesh connected WSN. From this, a suitable WSN design could be chosen so that it would be realized to work efficiently.

The structure of the paper is as follows. Section 2 depicts the modeling of the actual light traps network. This is a followed by an analysis of light traps wireless sensor networks as graphs and simulations, in section 3. Next section describes some experimental characteristics resulting from application of radio communication into the light traps network in Hau Giang province in Mekong Delta. All the results are simulated in Occam [16] programs automatically produced from NetGen [3]. Section 5 interprets the light traps WSN as a cyber physical system. In the last section we draw some conclusion regarding the presented work in particular, as well as possible future approaches in general.

## II. MODELING THE LIGHT TRAPS SURVEILANCE NETWORK

BPH is an incomplete metamorphosis insect. Its life cycle is composed of three stages: egg, nymph, and adult. Depending on the food source and the population density, BPH can dispatch following the wind direction on a very large scale [22]. During the migration process, the BPHs can reproduce themselves to create the next generation of hoppers [23].

A light trap uses light as an attraction source. Light traps depend on the positive phototactic response of the insects, physiological as well as abiotic environmental factors can influence the behavior [13]. The light trap effectively controls the major pests infesting rice, specifically green leafhopper, zigzag leafhopper, white-backed planthopper, brown planthopper, rice whorl maggot, pink stemborer, and several unidentified species of leaf folders and stem borers [21]. Light traps have been set up all over the Mekong Delta. The light is usually turned on at 7:00pm everyday and the sample is collected and analyzed in the next morning [14].

Light traps are effective in monitoring insects. Firstly, light traps can kill insects when they are attracted. In Mekong Delta, a light trap may destroy millions of BPH per night in the peak time of BPH [10]. Secondly, people can make decisions relating to their fields thanks to monitoring data in light traps. For instance, people can choose the time to spray insecticides, the time to sow, and make other decisions.

*Research supported by DREAM Team/UMI 209 UMMISCO-IRD, Can Tho University.

Bao Hoai Lam is with the Can Tho University, Vietnam (email: lhbao@cit.ctu.edu.vn).
Tai Tan Phan is with the Can Tho University, Vietnam (email: pttai@cit.ctu.edu.vn).
Long Huynh Vuong is with the Tay Do University, Vietnam (email: vuonghuynhlong@gmail.com).
Hiep Xuan Huynh is with DREAM Team/UMI 209 UMMISCO-IRD, Can Tho University, Vietnam (email: hxhiep@cit.ctu.edu.vn).
Bernard Pottier is with the Université de Bretagne Occidentale, LAB-STICC, UMR CNRS 6285, France (email: pottier@univ-brest.fr).

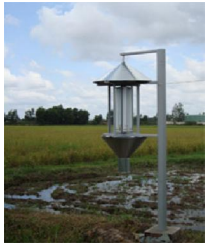

Figure 1.  A light trap [15]

An abstraction for a light traps network is a graph G = (V, E), where V is the set of vertices and E is the set of edges. In a light traps network, the set V is equivalent with the set of light traps, V = {$v_1$, $v_2$,... , $v_n$}. If the Euclidean distance of two light traps $v_i$ and $v_j$ is at most a given radius r (d($v_i$, $v_j$) ≤ r),  an edge between $v_i$ to $v_j$ will be  added into E [1].

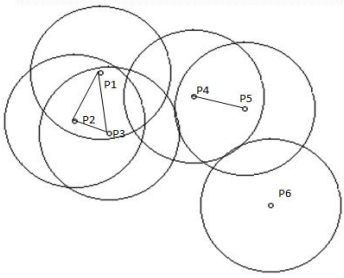

Figure 2.  A light traps network as a graph

Figure 2 illustrates the logical graph of a light traps network where the dots mean the vertices in V and the lines mean the edges in E. The graph contains 5 nodes with 2 sub graphs and an isolated node. The sub graph 1 consists of 3 nodes P1, P2, P3 since distances among these nodes are smaller than the given radius r. Similarly, P4 and P5 become another sub graph while the node P6 is isolated.

As the Euclidean distance of two light traps is at most the radius r, a light trap is *under the influence* of another. This influence is the *estimated* BPH ability to migrate from one to another.  Actually, BPH can migrate very far from the source point. Some migrations of 700 km have been identified by mark and recapture experiments in China [20]. This ability of migration depends on temperature, pressure, rain, and mostly on wind [19].

The above radius r is called the maximum distance which BPH can dispatch *following* the wind. Suppose that v is the wind velocity and t is the sampling time per day (normally 4 hours/day from 7:00pm to 11:00pm), then, the maximum distance which BPH can dispatch is: r = v * t [22].

Support that:

- $N_1(x_1, y_1)$ is the coordinate of the light trap $v_1$. $N_1$ is also called a source point from which BPH can dispatch.

- $N_2(x_2, y_2)$ is the coordinate of the light trap $v_2$.

- $D_1(x_1', y_1')$ is the farthest point to which BPH can reach toward the wind direction.

- β is the angle made by the $N_1N_2$ line and the $N_1D_1$ line.

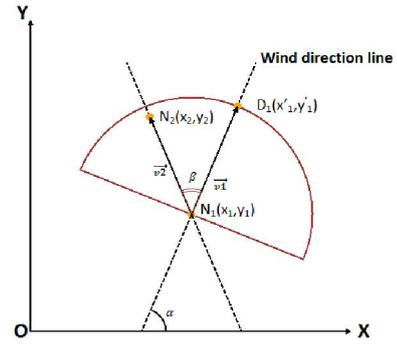

Figure 3.  Two light traps in a plane

Then, an edge is established between $N_1$ and $N_2$ if and only if:

$$\begin{cases} |N_1N_2| \leq v*t \\ |\beta| \leq \dfrac{\pi}{2} \end{cases} \quad [22]$$

Example when the wind velocity is 2km/h and the sampling time is 4h, the radius should be r = 8km. Besides, some markers in figure 4 are light trap positions. In this case, the graph for the light trap network is shown as below:

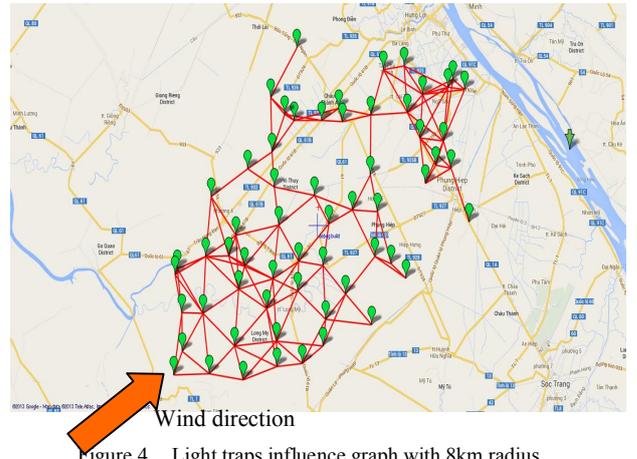

Figure 4.  Light traps influence graph with 8km radius

## III.  WIRELESS SENSOR NETWORKS

A WSN consists of n wireless sensor nodes distributed on a surface [2] (figure 5). Each wireless sensor node has an omni-directional antenna, allowing a transmission from a node to be received by all nodes in its vicinity. The radius of this node is called the *transmission range* of the sensor. Besides, we assume that all nodes have the same transmission range. These wireless sensor nodes define a graph in which there is an edge between 2 nodes *u, v* if only if their Euclidean distance is at most the transmission range (d(*u, v*) ≤ range) [2, 20].

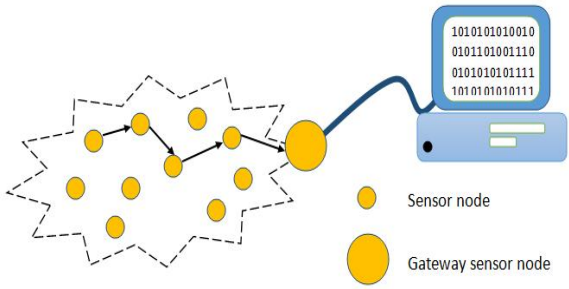

Figure 5. A wireless sensor network

In a WSN, two nodes within each other's transmission range can communicate directly while too distant nodes can communicate through multi-hop wireless links using intermediate nodes to relay the message. All nodes within a constant $k$ hops of node $u$ in the network are called a $k$-hop neighbors of $u$, or node $u$ has the diameter k [2]. The diameter k plays an important role in routing data packets to the gateway and controlling algorithms.

To implement this, routing tables allow each node to know what is the best link to reach a particular destination node. A known algorithm to do this after a connectivity setup, is to repetitively accept and send a working routing table from, and to direct neighbors. In the initial state, each node set up its *Identity* in an entry of the node with a distance of zero. Then, each participant must execute a loop at least diameter times.

- For each link, each incoming table is analyzed to find new discovered nodes.
- These nodes are added to the working table with their distance increased by one.
- The index of the link is kept in the discovered node entry.

At the end of diameter loops, each node has received the identity of each other node in its network, it also has the index of the link to use to join this other node in the shortest way. Therefore, we have a routing structure usable to send data packets to destinations known by their *Identity*. By adding an accurate buffering system avoiding contention, we can add a transport system and reach a gateway from anywhere. This is enough to cover practically the needs of data collection.

## IV. EXPERIMENTS

### A. Preparing data

Real light trap positions in 2011 in Hau Giang are used in this experiment. The first step is to locate all light trap positions in the map. Each marker in figure 4 is also the position of a light trap in Hau Giang (60 light traps).

When considering all light traps as a WSN, each light trap position is a sensor position. Thus, the WSN has 60 sensor nodes that also include a gateway.

### B. Light traps surveillance network as a WSN

NetGen [3] allows modeling and exploring distributed behaviors from practical deployment situations [7, 8, 9]. The first step in modeling is to obtain an abstract representation of a sensor distribution. In the case of light traps in Hau Giang, positions and distance between traps will condition communications together with the radio technology, and the power applied to radio emitters that define the transmission range. In practice, this design step is based on a map where sensors are placed, in relation with a distance metric. The radio range can be tuned in the graphic interface that reacts by drawing the corresponding connectivity (slider Figure 4).

In this case, it appeared that ranges from 5km to 40km were interesting, therefore, it was decided to study nine networks with 5, 6, 7, 8, 9, 10, 20, 30, and 40 km ranges. The remarkable point is that the resulting networks will lead to very different situations such as:

- Disrupted network.
- Single network with large diameter (large number of hops to reach any node from any other node).
- Over-connected network with too much communication links leading to transmission conflicts.

Some examples of these resulting networks are shown as follow:

*1) 5km range network (disrupted network)*

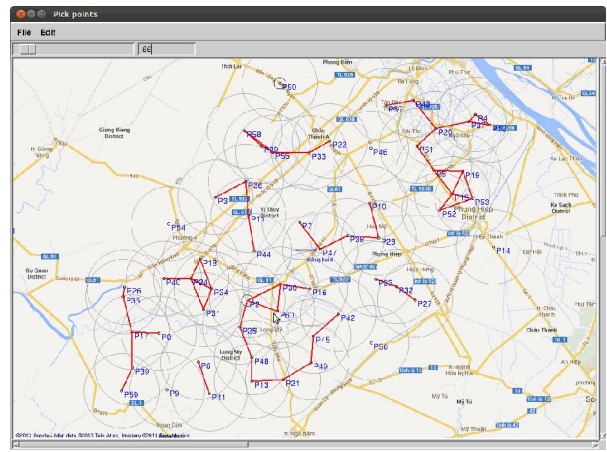

Figure 6. 5km range WSN showing uncomplete connectivity

The logical network is shown below:

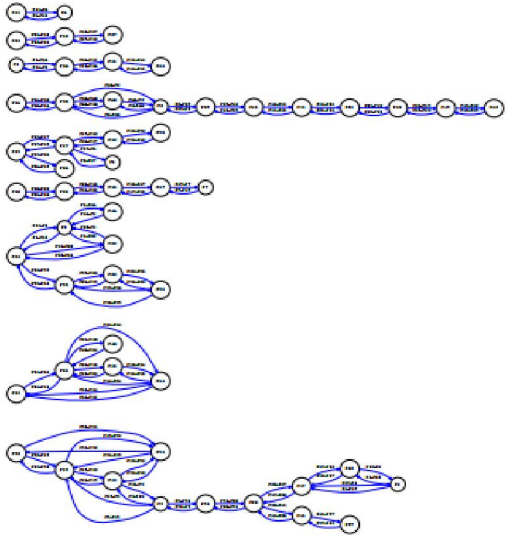

Figure 7.  Equivalent logical network given a range for 5km (9 sub networks)

*2) 8km range network (single network, large diameter)*

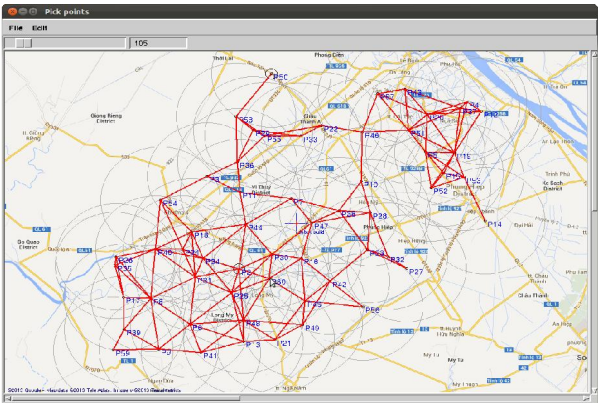

Figure 8.  A WSN with 8km in NetGen

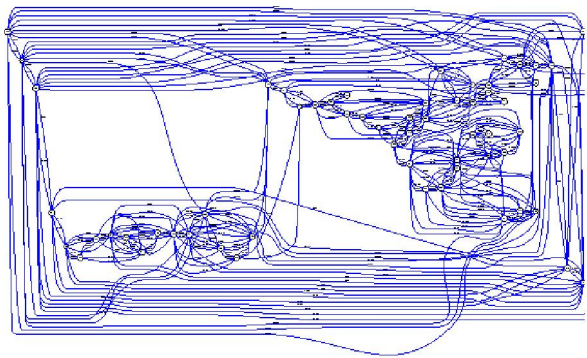

Figure 9.  Equivalent logical network given a range for 8km

Figure 9 also illustrates the layout for the 8km WSN (figure 8). Each circle represents a zone controlled by a radio transceiver while edges represent communication links.

NetGen tool also allows generating the logic representation of the 60 nodes network (Figures 7 and 9).

*3) 20 km network (over-connected network)*

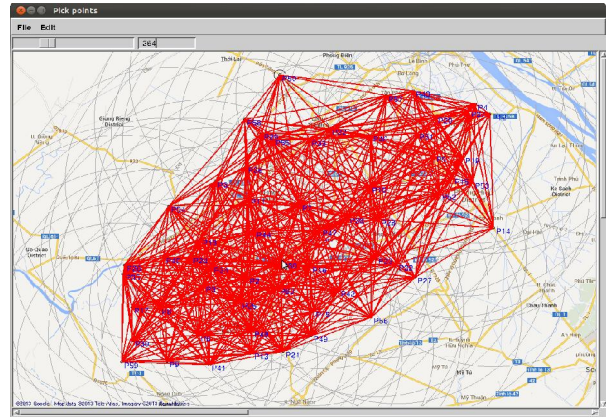

Figure 10.  20 km WSN

C. *Some simulation results*

Table I depicts the number of sensor nodes in different radio ranges. From 5km to 7km, the number of bound nodes is 55, 57, and 58, respectively. This means that there remain some isolated nodes (5, 3, and 2, respectively). From 8km to 40km, networks contain all the 60 nodes. These networks could be suitable for the Hau Giang control and data collection.

| Ranges (km) | Nodes | Channels | Min fan-out | Max fan-out |
|---|---|---|---|---|
| 5 | 55 | 110 | 1 | 4 |
| 6 | 57 | 168 | 1 | 5 |
| 7 | 58 | 222 | 1 | 7 |
| 8 | 60 | 298 | 1 | 9 |
| 9 | 60 | 376 | 2 | 10 |
| 10 | 60 | 476 | 3 | 13 |
| 20 | 60 | 1548 | 13 | 39 |
| 30 | 60 | 2566 | 26 | 58 |
| 40 | 60 | 3192 | 39 | 59 |

TABLE I.  NUMBER OF INTEGRATED NODES, NUMBER OF LINKS, FAN-OUT MIN AND MAX, FOR EACH CONSIDERED RANGE

Table I also describes the number of channels and the fan-out information of WSN with various ranges. Starting with 110 communication channels in the 5km network (or simply edges in the logical network in Figure 7), the number of channels has almost linear rise with 3192 channels in range 40km. Similarly, the minimum and maximum fan-out in the networks raise slightly when the range rises (large range produces excessive connectivity). Therefore, WSN become more complex (and inefficient) when the range increases.

To run a simulation, 2 Occam [16], or CUDA [6] program parts are assembled:

1. **Architecture**: automatically generated in Occam syntax from NetGen tools to implement the process graph (it is almost impossible to do it by hand, see Table I).

2. **Behavior**: set of procedures representing the nodes at work, especially their contribution to distributed control,

routing and data collection tasks. These procedures repetitively execute phases for the synchronous model [5]:

- Produce messages in output buffers to prepare communications
- Send and receive in parallel on node links (this figures transmit and receive radio operations)
- Analyze input buffers contents, and integrate local data in local status.
- Sleep, and wait the newt synchronous step.

An example program is to produce metrics as it will be computed dynamically in the final network. The trace output is the *diameter* and a leader *Identity*. Diameter will then be used to reduce the number of steps in algorithms to the useful minimum. *Leader Identity* are useful for control algorithms, as an example to distinguish a gateway.

When running this basis behavior on Hau Giang networks, we have obtained the following result:

| Range (Km) | Networks | Diameter max |
|---|---|---|
| 5 | 9 | 9 |
| 6 | 2 | 15 |
| 7 | 1 | 15 |
| 8 | 1 | 12 |
| 9 | 1 | 9 |
| 10 | 1 | 7 |
| 20 | 1 | 4 |
| 30 | 1 | 3 |
| 40 | 1 | 2 |

TABLE II. NUMBER OF NETWORKS AND DIAMETERS

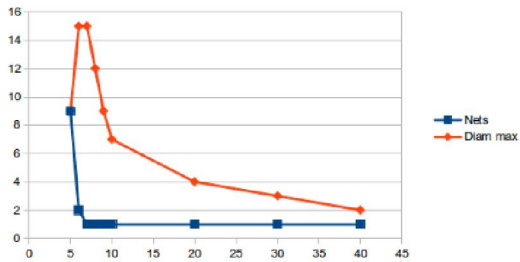

Figure 11. Graph for number of networks and their maximum diameter

Figure 11 depicts the number of networks (sub network) of WSN with ranges from 5, to 40 km. Starting with 9 sub networks in range 5km (figure 7), the number of networks decreases dramatically with 1 network in range 7km. At 8km, all nodes are integrated, the number of networks remaining 1 until 40km.

Some of these numbers are also indicative of insufficient connectivity (5km and 6km cases), or overcrowd connectivity that will lead to lot of collisions if nothing is done to schedule communication. Nevertheless, when the range increases, the transmission range is large enough to cover all sensor nodes leading to a single well balanced network.

Figure 11 also illustrates the maximum diameter of WSN with different ranges. In 5km range network, the maximum diameter is 9, then, suddenly it rises to 15 hops in 6 and 7 km range network. From 15 hops in 7km range, the maximum diameter smoothly decreases to 2 in 40km range. The reason is also in the transmission range. As it becomes larger, there are more links among sensor nodes. Thus, a node in these networks has a larger neighborhood than a node in small range networks.

For the above analysis in Table I and Figure 11, the 8km network could be the best choice for data collection in Hau Giang. This network enables to test algorithms, delays, latencies, bandwidths. It needs 12 synchronous steps to propagate data everywhere, and particularly to a gateway sink.

*D. Routing to the gateway*

Table III depicts number of hops to reach a possible leader/gateway node for different ranges. The depth leader measure is the minimum number of synchronous steps (hops) necessary to bring all data to an arbitrary gateway.

| Range | Depth leader |
|---|---|
| 5 | 8 |
| 6 | 8 |
| 7 | 12 |
| 8 | 9 |
| 9 | 6 |
| 10 | 5 |
| 20 | 2 |
| 30 | 2 |
| 40 | 1 |

TABLE III. DEPTH LEADERS

In the case of the 8km network (Figure 8, 9), a leader gateway has been set up to be the P43 node with Id 60. There is no isolated node.

At a distance of 1, the syntax of abstract specification of the networks displays 4 neighbor nodes as follows:

*P43 { P20, P37, P51, P57 } Node (703 @ 74) (106)*

The sensor will execute the *Node* program, and is located at coordinates (x=703, y=74) on the map.

Table III illustrates that there are at least 9 synchronous steps to bring all data to a P43 gateway. Real data transmission is done by radio emitting packets. These packets are possibly delayed in routing nodes, and propagated toward the gateway: if a link bandwidth is saturated, it is necessary to keep data in FIFO, waiting for less activity. These conditions can be detected and managed during the development cycle. If the WSN frequency is tuned to a 5 minute sleep time, 9 steps could mean a (9 x 5 = 45 minutes) latency before distant information can reach the gateway.

V. THE BPH SURVEILLANCE NETWORK - A CPS SYSTEM

A Cyber Physical System (CPS) is a system of collaborating computation controlling physical processes. Embedded computers and networks monitor and control the physical processes, usually with feedback loops where physical processes affect computations and vice versa [18]. In physical world, the passage of time and concurrency are two core characteristics [18].

The BPH surveillance network in the Mekong Delta fits into a CPS framework. BPH, rice, weather become physical entities while the observation network is the computation.

Light traps in the network can destroy a massive amount of BPH. Besides, thanks to monitoring the densities of BPH in light traps, people can have some decisions relating to their fields. These build a physical loop between physical entities and computation with timed characteristics.

## VI. CONCLUSION AND FINAL REMARKS

In this paper, a light traps surveillance distribution has been modeled and investigated as a WSN. In our model, each light trap is considered as a sensor with a radio link. The influence radius of a light trap becomes the transmission range of a sensor node. The topology graph is computed, based on the influence radius close to the radio range.

We simulate several WSN options for light traps network in Hau Giang province in the Mekong Delta using NetGen. From the simulation results, useful networks are those from 8km to 40km range. The 8km range connects all traps in a mesh being a capable choice for the Hau Giang case. This network can use *CSMA/CA* synchronous periodic exchanges between nodes. We know how to dimension data structures, packets, and algorithms for an actual implementation.

Instead of using mesh routing, it is possible to switch more powerful radio emitting and reach 20km ranges. In this case, the number of hops is smaller, but the number of collision will increase, possibly prevented by star topology and *time division* control style. Switching dynamically is possible.

*Dedicated sensors* inside light traps could allow classifying insects, given that a light trap attracts different kinds of insect and each insect has different color and shape.

When large swarms (*clouds*) appear, the trap can destroy a massive amount of BPH (some kilograms !), in these cases sensing the weight of collected BPH is an alternative to the identification of insects.

The *coverage* problem, common in WSN domain, need to be clarified. From a farmer perspective, it appears as a very local *insect density measure*. The light traps are useful for people in making decisions relating to their rice crops. However, in practice, a light trap *network* is expected to inform about thousands of hectare of rice area by deducing migrations from the value of these measures. *Real time, distributed information samples* will be interesting to guess location, direction, and densities of insect groups.

BPH migrations can appear as *clouds*, and the light traps may not be efficient enough to detect these migrations, the light having limited effects. Some other methods need to be investigated, such as radar remote sensing systems. These systems may also help monitoring real time BPH migration.

In this paper, the BPH surveillance network is considered as a cyber physical system. Process simulation, together with the synchronous model support precise knowledge of system and physical behaviors. Probably this way of working allows hiding hardware details at the profit of functions that contribute to the sensing engine. However, we need to take into account some theories that can blend physical processes and computation. Hybrid systems may be capable.